\documentclass[journal]{IEEEtran}
\newcommand{\ee}{\epsilon}			
\usepackage{epsf,psfrag,amssymb,amsfonts,color,cite,fancybox}
\usepackage[mathscr]{eucal}
\usepackage[dvips]{graphicx}
\usepackage{ifpdf}
\usepackage{longtable}	
\usepackage{subfigure}
\usepackage{multicol}
\ifCLASSINFOpdf
\else
\fi
\usepackage[cmex10]{amsmath}
\begin{document}
%
\title{Some issues with Quasi-Steady State Model in Long-term Stability}
%
%
%

\author{Xiaozhe Wang,  
        Hsiao-Dong Chiang,~\IEEEmembership{Fellow.}
\thanks{Xiaozhe Wang is with the Department
of Electrical and Computer Engineering, Cornell University, Ithaca,
NY, 14853 USA e-mail: xw264@cornell.edu}
\thanks{Hsiao-Dong Chiang is with the Department of Electrical and Computer Engineering, Cornell University, Ithaca, NY 14853 USA email:hc63@cornell.edu}
}

\maketitle

\begin{abstract}
The Quasi Steady-State (QSS) model of long-term dynamics relies on the idea of time-scale decomposition. Assuming that the fast variables are infinitely fast and are stable in the long-term, the QSS model replaces the differential equations of transient dynamics by their equilibrium equations to reduce complexity and increase computation efficiency. Although the idea of QSS model is intuitive, its theoretical foundation has not yet been developed. 
In this paper, several counter examples in which the QSS model fails to provide a correct approximation of the complete dynamic model in power system are presented and the reasons of the failure are explained from the viewpoint of nonlinear analysis.
\end{abstract}


%
\IEEEpeerreviewmaketitle

\section{Introduction}
\IEEEPARstart{T}he ever-increasing loading of transmission networks together with a steady increase in load demands has pushed the operation conditions of many power systems ever closer to their stability limits \cite{Chiang:book}-\cite{SauerPai:book}. Voltage stability has become one of the major concerns for the secure operation of power systems. Voltage stabilities are classified into transient voltage stability, mid-term stability and long-term stability based on different time scales. The distinction between mid-term and long-term can be based on neither fixed time-frame basis nor modelling requirements\cite{Kundur:book}, hence we only use long-term time scale in this paper to denote the one beyond the transient time scale for stability analysis. This paper considers the long-term voltage stability model.

Power system dynamic models are large and involve different time scales, and it is time-consuming and data-demanding to simulate the dynamic behaviors over long time intervals. Based on the idea of time scale decomposition, the quasi steady-state (QSS) \cite{Cutsem:book}\cite{Cutsem:artical} seeks to reach a good compromise between accuracy and efficiency. However, there are certain limitations of the QSS model such as singularity problem. When this happens, the Newton iterations diverge in practice and the simulation cannot proceed. There are several papers that addressed the singularity problem and tried to solve it by a combination of detailed simulation and the QSS approximation\cite{Cutsem:artical2}, Newton method with optimal multiplier \cite{Cutsem:artical3}, and continuation method \cite{Wang:artical}.

However, less attention has been paid to a severe situation when the assessment based on the QSS model is not reliable. In this situation, the QSS model gives incorrect stability assessments in long-term stability analysis which means the QSS model concludes the stability of the complete model, which is in fact unstable. Due to the existence of these situations, the QSS model may not consistently give conservative stability analysis of the complete model. In other words, the QSS model does not work under certain conditions, thus sufficient conditions are needed under which the QSS model provides correct stability assessment of the complete model.

This paper is organized as follows. Section \ref{sectiondymodel} and Section \ref{sectionqssmodel} briefly introduce the basic concept of complete dynamic model and the QSS model with numerical examples. Section \ref{counterexample} presents two counter examples in which the QSS model fails to provide correct approximations of the complete model. Specifically, while the QSS model is stable, the complete model suffers from voltage instabilities. Also, theoretical explanation for this failure is presented. Conclusions and perspectives are stated in Section \ref{conclusion}.
\section{Complete Dynamic Model}\label{sectiondymodel}

The complete power system model for calculating system dynamic response relative to a disturbance comprises a set of first-order differential equations and a set of algebraic equations\cite{Chiang:book}-\cite{SauerPai:book}. The algebraic equations:
\begin{equation}
{0}={g}({z_c,z_d,x,y})\label{algebraic eqn}
\end{equation}
describing the electrical transmission system and the internal static behaviors of passive devices. While the transient dynamics are captured by differential equations:
\begin{equation}
\dot{{x}}={f}({z_c,z_d,x,y})\label{longtermeqn}
\end{equation}
which describe the internal dynamics of devices such as synchronous generator and its associated excitation system, interconnecting transmission network together with static load, induction and synchronous motor loads, as well as other devices such as HVDC converter and SVC. ${f}$ and ${g}$ are smooth functions, and vectors ${x}$ and ${y}$ are the corresponding short-term state variables and algebraic variables respectively. Both continuous equations and discrete-time equations are needed to represent long-term dynamics:
\begin{eqnarray}\label{longtermcontinuous}
\dot{z}_{c}&=&\ee{h}_c({z_c,z_d,x,y})\\
{z}_d(k+1)&=&{h}_d({z_c,z_d(k),x,y})\label{longtermdiscrete}
\end{eqnarray}
where ${z}_c$ and ${z}_d$ are the continuous and discrete long-term state variables respectively, and $1/\ee$ is the maximum time constant among devices. These equations describe the dynamics of exponential recovery load and thermostatically recovery load, turbine governor, LTC, OXL and armature current limiter, as well as shunt capacitor/reactor switching all belongs to long-term dynamics. Note that shunt switching and LTC are typical discrete components captured by Eqn (\ref{longtermdiscrete}).

Usually, transient (model) dynamics have much smaller time constants compared with those of long-term dynamics, as a result, $z_c$ and $z_d$ are also termed as slow state variables, and $x$ are termed as fast state variables. If we represent the above equations in $\tau$ time scale where $\tau=t\ee$, and we denote $\prime$ as $\frac{d}{d\tau}$, then we have:
\begin{eqnarray}
\ee{{x}}^\prime&=&{f}({z_c,z_d,x,y})\\
{z}_{c}^\prime&=&{h}_c({z_c,z_d,x,y})\\
{z}_d(k+1)&=&{h}_d({z_c,z_d(k),x,y})
\end{eqnarray}
Hence, the complete power system dynamic model involves different time scales which makes the time domain simulation over long time intervals very demanding. The QSS model based on time-scale decomposition is proposed in \cite{Cutsem:book}\cite{Cutsem:artical}\cite{Cutsem:artical4} and will be briefly stated in the following Section. 

\section{Quasi Steady-State Model}\label{sectionqssmodel}
The Quasi Steady-State (QSS) model is derived using the idea of time-scale decomposition and aims to offer a good compromise between the efficiency and accuracy\cite{Cutsem:artical}. In the QSS model, the differential equations describing transient dynamics are replaced by their equilibrium equations under the assumption that transient dynamics are stable and settle down infinitely fast in the long-term time scale.

Table \ref{table1} illustrates the concept of time-scale decomposition. The transient model is obtained by assuming that slow variables ${z}_c$ and ${z}_d$ are constant parameters. While in the QSS model, the transient dynamic equations (\ref{longtermeqn}) are replaced by the corresponding equilibrium equations:
\begin{equation}
{f}({z_c,z_d,x,y})=0
\end{equation}

\begin{center}
\begin{table}[h]
\centering
\caption{The mathematical description of model for power system}\label{table1}
\begin{tabular}{|c|c|}
\hline
complete model&
${z}_{c}^\prime={h}_c({z_c,z_d,x,y})$\\
&${z}_d(k+1)={h}_d({z_c,z_d(k),x,y})$\\
&$\ee{x}^\prime={f}({z_c,z_d,x,y})$\\
&${0}={g}({z_c,z_d,x,y})$\\
\hline
transient model&$\dot{{x}}={f}({z_c,z_d,x,y})$\\
(approximation for transient stability)&${0}={g}({z_c,z_d,x,y})$\\
short-term:0-30s&\\
\hline
QSS model
&${z}_{c}^\prime={h}_c({z_c,z_d,x,y})$\\
(approximation for long-term stability)&${z}_d(k+1)={h}_d({z_c,z_d(k),x,y})$\\
long-term:30s-a few minutes&${0}={f}({z_c,z_d,x,y})$\\
&${0}={g}({z_c,z_d,x,y})$\\
\hline
\end{tabular}
\end{table}
\end{center}

Under certain conditions, the QSS model performs quite well with similar accuracy as the detailed complete model, while it takes much less time to simulate if a larger time step or adaptive time steps are implemented. Also, compared with the complete model, the Jacobian matrix of the QSS model does not need to be updated at every time step, and it can be updated only following discrete events such as LTC or OXL activation unless slow convergence rate is observed\cite{Cutsem:book}. As a result, the QSS model is faster to simulate than the complete model. This paper focus on the accuracy and reliability of the QSS model instead of efficiency, thus the same time step as that of the complete model will be used and the Jacobian of the QSS model is updated at every time step as the complete model.

A numerical example is presented below which shows the trajectory comparison between the complete model and the QSS model. The QSS model and the complete model finally settle down to the same long-term stable equilibrium point (SEP) in this case, and the QSS model provides a good approximation of the complete model in long-term stability analysis. However this is not always true which can be seen from the counter examples presented in Section \ref{counterexample}.

\begin{figure}[!ht]
\centering
\begin{minipage}[t]{0.51\linewidth}
\includegraphics[width=1.95in ,keepaspectratio=true,angle=0]{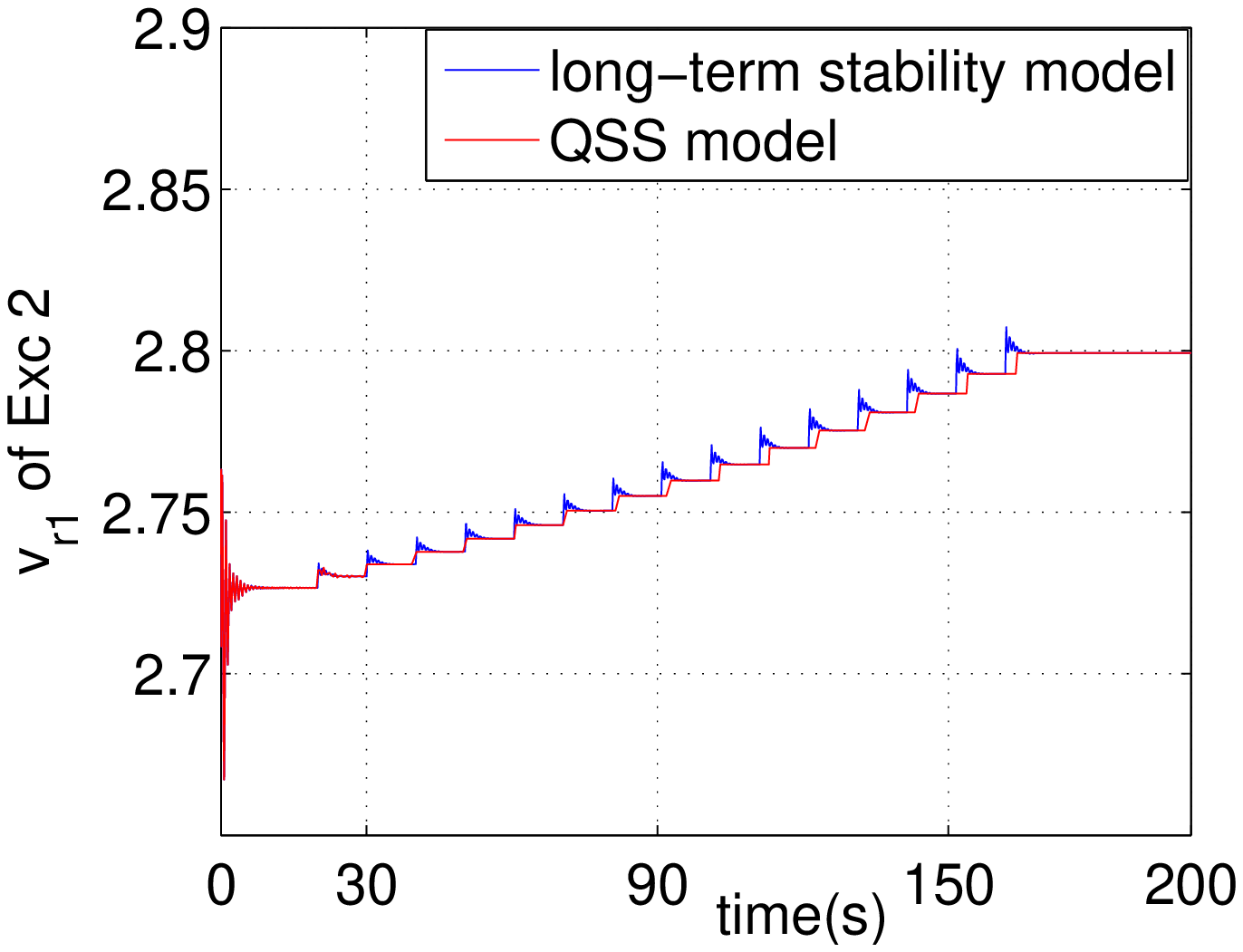}
\end{minipage}%
\begin{minipage}[t]{0.51\linewidth}
\includegraphics[width=1.95in ,keepaspectratio=true,angle=0]{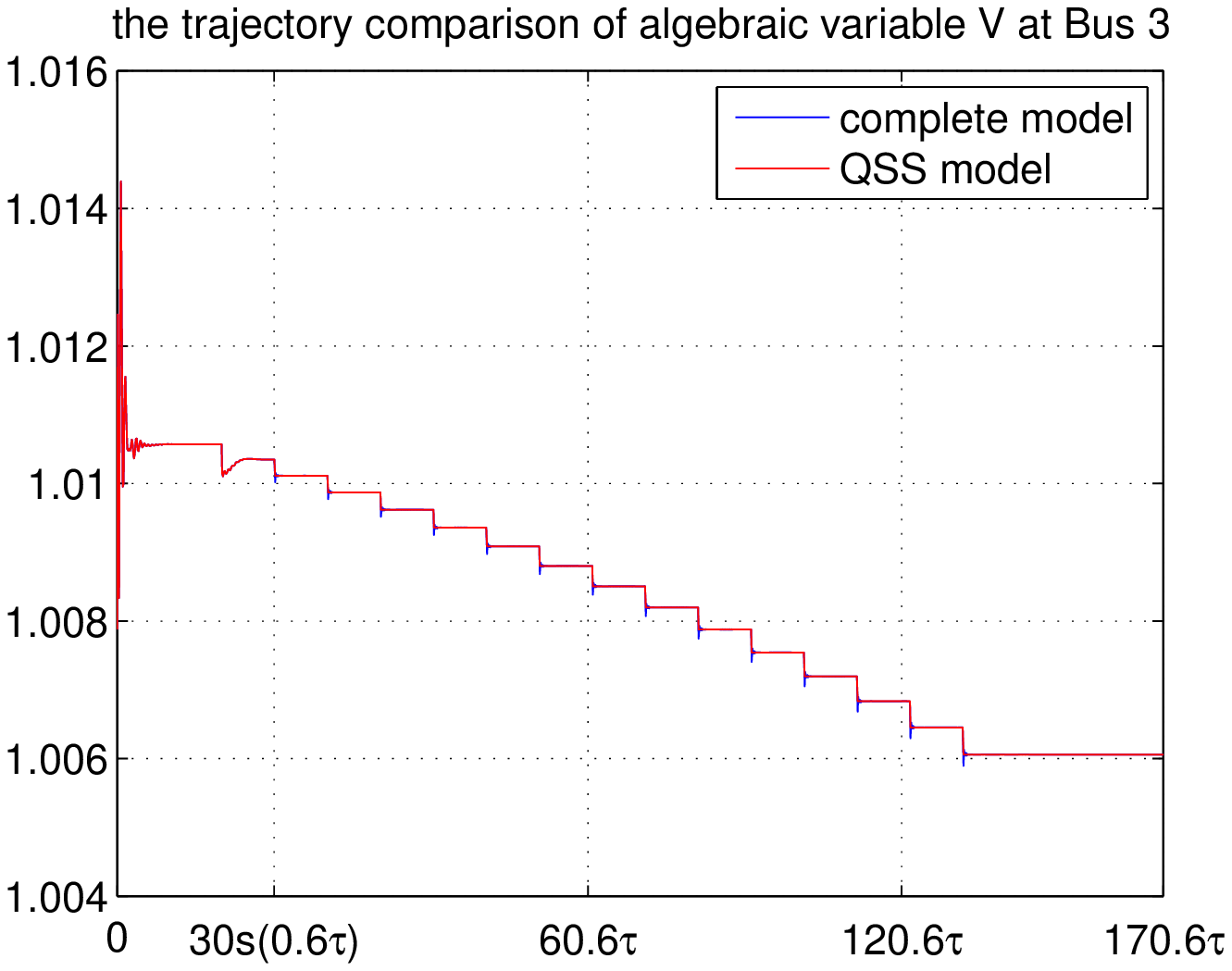}
\end{minipage}
\caption{The trajectory comparisons of the complete model and the QSS model for different variables. The trajectory of complete model followed that of the QSS model until both of them converged to the same long-term SEP.}\label{my14completeqss}
\end{figure}

The numerical study was performed using PSAT 2.1.6\cite{Milano:article} on a modified model of IEEE 14-bus test system whose one-line diagram is attached in Appendix \ref{numerical_ap}. There was a fault at Bus 9 at 1s and the fault was cleared at 1.083s by opening the breaker between Bus 10 and Bus 9. In the complete model, the fast variables settled down by 30s after the contingency, while the dynamics of load tap changer, turbine governors and exponential load evolved in a longer time period. The QSS model was used starting from 30s when transient dynamics almost settled down. When the QSS model was used, fast variables converged infinitely fast when slow variables evolved. Finally, both the QSS model and the complete model converged to the same long-term SEP. The comparison of trajectories of the complete model and the QSS model are shown in Fig. \ref{my14completeqss}.

\section{Nonlinear Framework: Stability Region}
Before presenting numerical examples, relevant definitions are needed to give a theoretical explanation of the simulation results. If we are interested in the study region $U_c={D_{z_c}}\times{D_{z_d}}\times{D_{x}}\times{D_{y}}$, both models have the same set of equilibrium points, that is $E=\{(z_c,z_d,x,y)\in{U}:z_d(k+1)=z_d(k),{h}_c({z_{c},z_{d},x,y})=0, {f}({z_c,z_d,x,y})=0,{g}({z_c,z_d,x,y})=0\}$. Assuming $(z_{cls},z_{dls},x_{ls},y_{ls})\in{E}$ is an asymptotically long-term SEP of both the QSS model and the complete model starting from $(z_{c0},z_{d0},x_0,y_0)$, and let $\phi_c(\tau,z_c,z_d,x,y)$ be the trajectory of the complete model and $\phi_q(\tau,z_c,z_d,x,y)$ be the trajectory of the QSS model starting from the same initial point, then the stability region for the complete model are defined as:
\begin{eqnarray}
&&A_c(z_{cls},z_{dls},x_{ls},y_{ls}):=\{(z_c,z_d,x,y)\in{U}:\phi_c(\tau,z_c,\nonumber\\
&&z_d,x,y)\rightarrow(z_{cls},z_{dls},x_{ls},y_{ls})\mbox{ as $\tau$}\rightarrow+\infty\}
\end{eqnarray}

For the QSS model, its dynamics are constrained to the set:
$\Gamma:=\{(z_c,z_d,x,y)\in{U}:{f}({z_c,z_d,x,y})=0, {g}({z_c,z_d,x,y})=0\}$
which is termed as the constraint manifold. Note that the constraint manifold may not be smooth due to the discrete behavior of $z_d$.
Then the stability region of $(z_{cls},z_{dls},x_{ls},y_{ls})$ for the QSS model are defined as:
\begin{eqnarray}
&&A_q(z_{cls},z_{dls},x_{ls},y_{ls}):=\{(z_c,z_d,x,y)\in{\Gamma}:\phi_q(\tau,z_c,\nonumber\\
&&z_d,x,y)\rightarrow(z_{cls},z_{dls},x_{ls},y_{ls})\mbox{ as $\tau$}\rightarrow+\infty\}
\end{eqnarray}

Similarly, for the transient model with fixing slow variables $z_c^\star$ and $z_d(k)$:
\begin{eqnarray}\label{transient model}
\dot{{x}}={f}({z_c^\star,z_d(k),x,y})\nonumber\\
{0}={g}({z_c^\star,z_d(k),x,y})
\end{eqnarray}
the equilibrium points are termed as transient SEPs. The stability region of transient SEP $(z_c^{\star},z_d(k),x_{ts},y_{ts})$ is defined as:
\begin{eqnarray}\label{transientsep}
&&A_t(z_c^{\star},z_d(k),x_{ts},y_{ts}):=\{(x,y)\in{D_x}\times{D_y},z_c=z_c^{\star},\nonumber\\
&&z_d=z_d(k):\phi_t(t,z_c^{\star},z_d(k),x,y)\rightarrow(z_c^{\star},z_d(k),\nonumber\\
&&x_{ts},y_{ts})\mbox{ as t}\rightarrow+\infty\}\nonumber\\
&&
\end{eqnarray}
where $\phi_t(t,z_c^{\star},z_d^{\star},x,y)$ is the trajectory of the transient model (\ref{transient model}). A comprehensive theory of stability regions can be found in \cite{Chiang:article1988}\cite{Chiang:article1989}\cite{Zaborszky:article}\cite{Alberto:article}.

Generally, the SEPs of each transient model are isolated and the trajectory $\phi_q(\tau,z_c,z_d,x,y)$ of the QSS model does not meet the singular surface and is constrained on $\Gamma_s$ all the time where $\Gamma_s$ is defined as:
\begin{eqnarray}\label{gammas}
&&\Gamma_s=\{(z_c,z_d,x,y)\in\Gamma: \mbox{all eigenvalues } \lambda \mbox{ of}(\frac{\partial{f}}{\partial{x}}-\frac{\partial{f}}{\partial{y}}\nonumber\\
&&{\frac{\partial{g}}{\partial{y}}}^{-1}\frac{\partial{g}}{\partial{x}})
\mbox{ satisfy Re}(\lambda)<0, \partial{g}/\partial{y}\mbox{ is nonsingular}\}\nonumber\\
&&
\end{eqnarray}
Note that each point of $\Gamma_s$ is a SEP of the transient model defined in Eqn (\ref{transient model}) for fixed $z_c^{\star}$ and $z_d(k)$. Thus generally the trajectory $\phi_q(\tau,z_c,z_d,x,y)$ of QSS model moved along $\Gamma_s$ on which each point is a SEP of the corresponding transient model. Given enough simulation time which is usually to be several minutes, both the QSS model and the complete model converge to the same long-term SEP.

\begin{figure}[!ht]
\centering
\includegraphics[width=3.5in,keepaspectratio=true,angle=0]{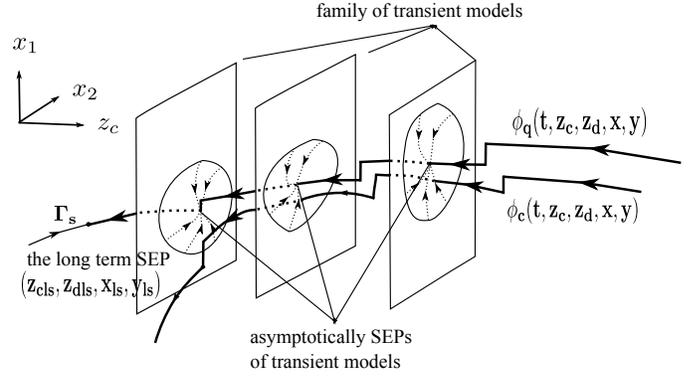}\caption{When $z_d$ firstly change to $z_d(k)$, the initial point of the complete model get outside of the stability region of the transient model and the trajectory of the complete model moves far way from the QSS model from then on.}\label{QSSzdunstable}
\end{figure}

However, if when $z_d$ firstly change from $z_d(k-1)$ to $z_d(k)$, and the initial point $(z_c^{\star},z_d(k),x_0,y_0)$ on the trajectory $\phi_c(\tau,z_c,z_d,x,y)$ lies outside the stability region $A_t(z_c^{\star},z_d(k),x_{ts},y_{ts})$ of the transient model:
\begin{eqnarray}\label{transient2}
\dot{x}&=&{f}({z_{c}^{\star},z_{d}(k),x,y})\\
{0}&=&{g}({z_{c}^{\star},z_{d}(k),x,y})\nonumber
\end{eqnarray}
then $\phi_c(\tau,z_c,z_d,x,y)$ will move away from the slow manifold $\Gamma_s$ as shown in Fig. \ref{QSSzdunstable}. As a result, the trajectory $\phi_c(\tau,z_c,z_d,x,y)$ of the complete model will not converge to the long-term SEP $(z_{cls},z_{dls},x_{ls},y_{ls})$ which the trajectory $\phi_q(\tau,z_c,z_d,x,y)$ of the QSS model converges to. Hence, the QSS model is not an appropriate approximation for the complete model and gives incorrect stability assessments in this case.

\section{Counter Examples}\label{counterexample}
The QSS model has some limitations in dealing with severe disturbances. As stated in \cite{Cutsem:book}, the QSS model cannot reproduce the instabilities where the slow variables trigger instability of fast variables. This means the QSS model can not capture the insecure cases when the fast variables are excited by the slow variables, thus result in voltage instabilities. In addition, the QSS model may converge to another stable equilibrium point different from the one the complete model converges to. Under these two situations, the QSS model does not capture the dynamic behavior of the complete model and give inaccurate approximations of the complete model. In brief, the QSS model can lead to incorrect stability assessment.


\subsection{Numerical Example I}\label{qss_numerical}
\begin{figure}[!ht]
\centering
\begin{minipage}[t]{0.5\linewidth}
\includegraphics[width=1.8in ,keepaspectratio=true,angle=0]{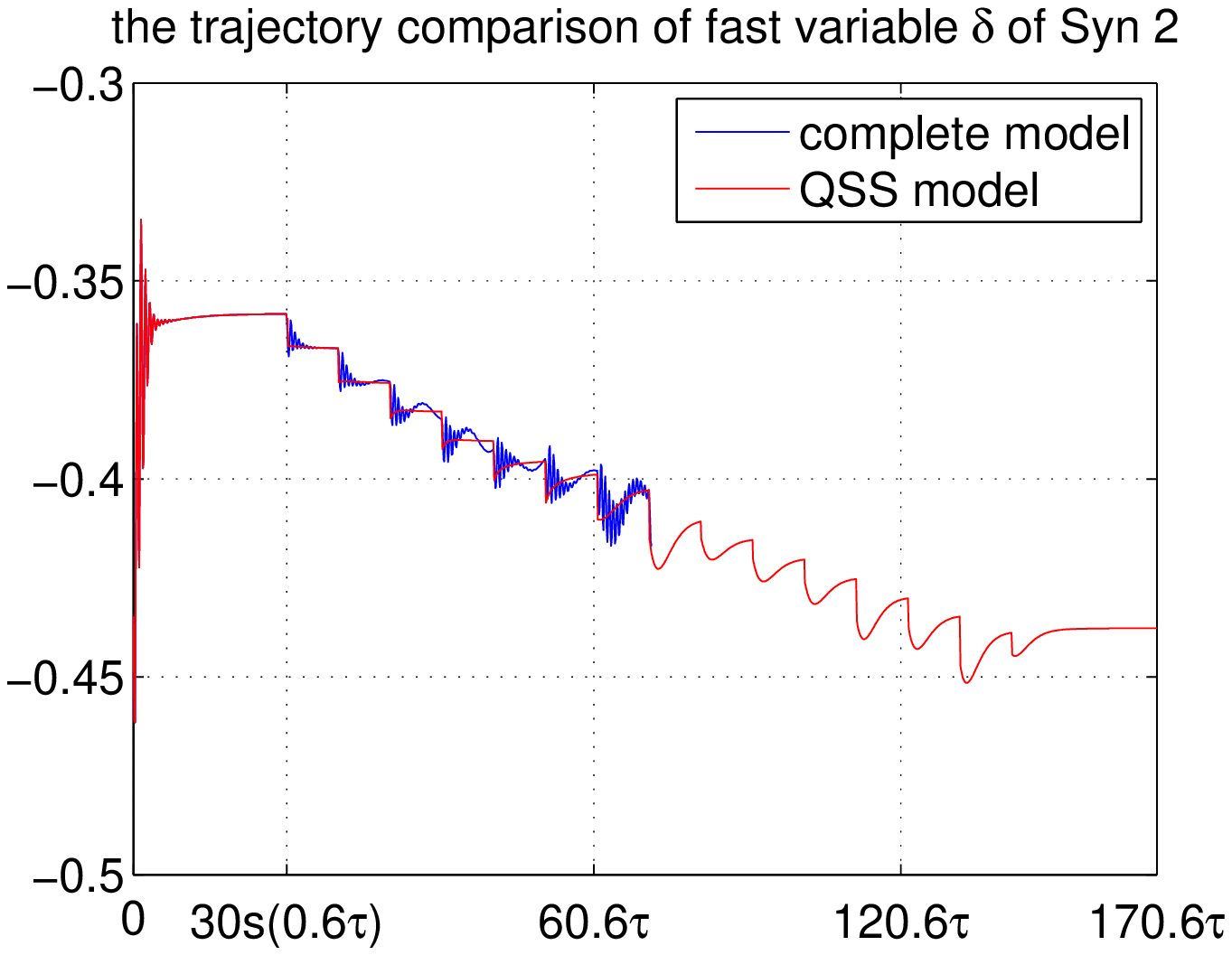}
\end{minipage}%
\begin{minipage}[t]{0.5\linewidth}
\includegraphics[width=1.8in ,keepaspectratio=true,angle=0]{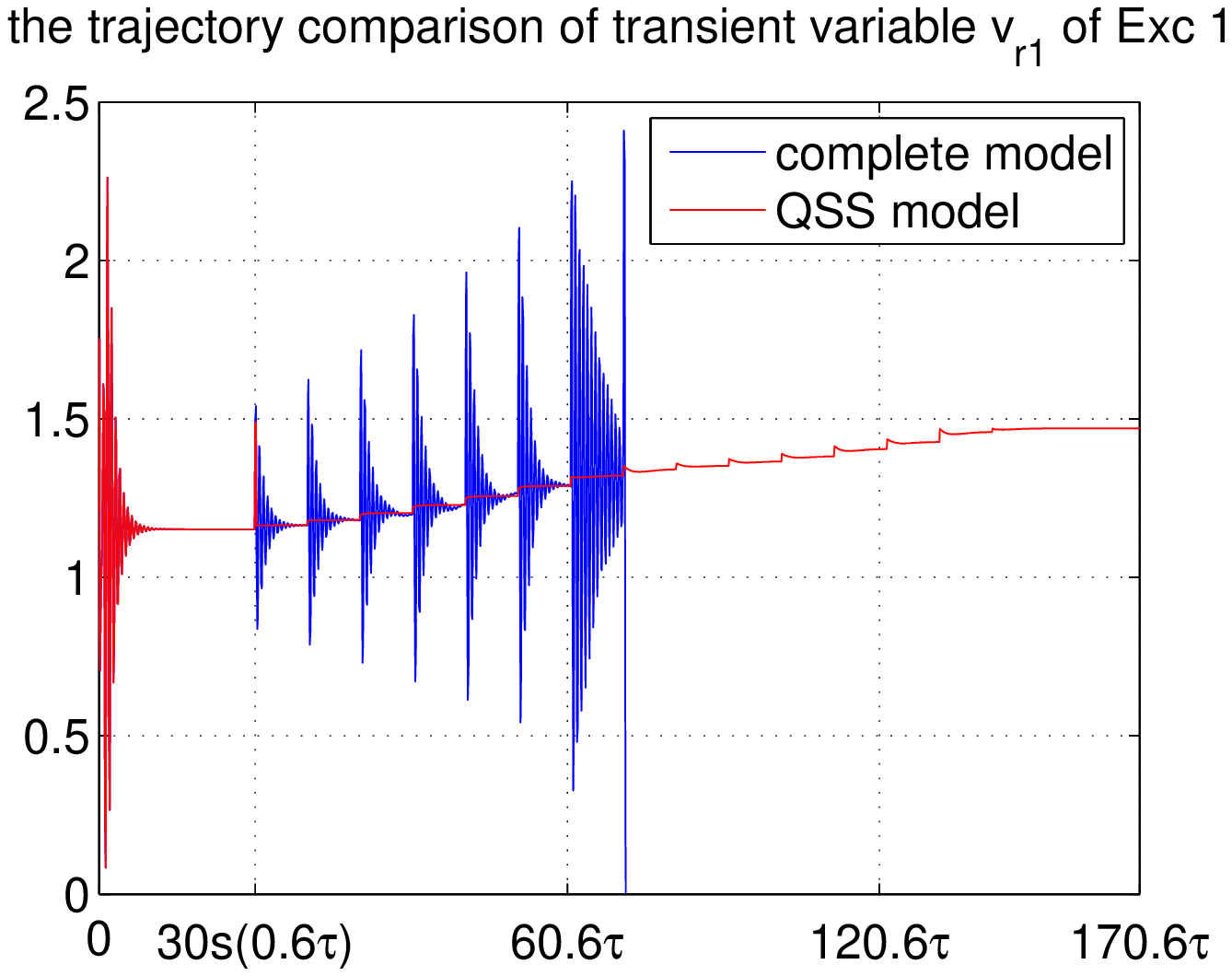}
\end{minipage}
\begin{minipage}[t]{0.5\linewidth}
\includegraphics[width=1.8in ,keepaspectratio=true,angle=0]{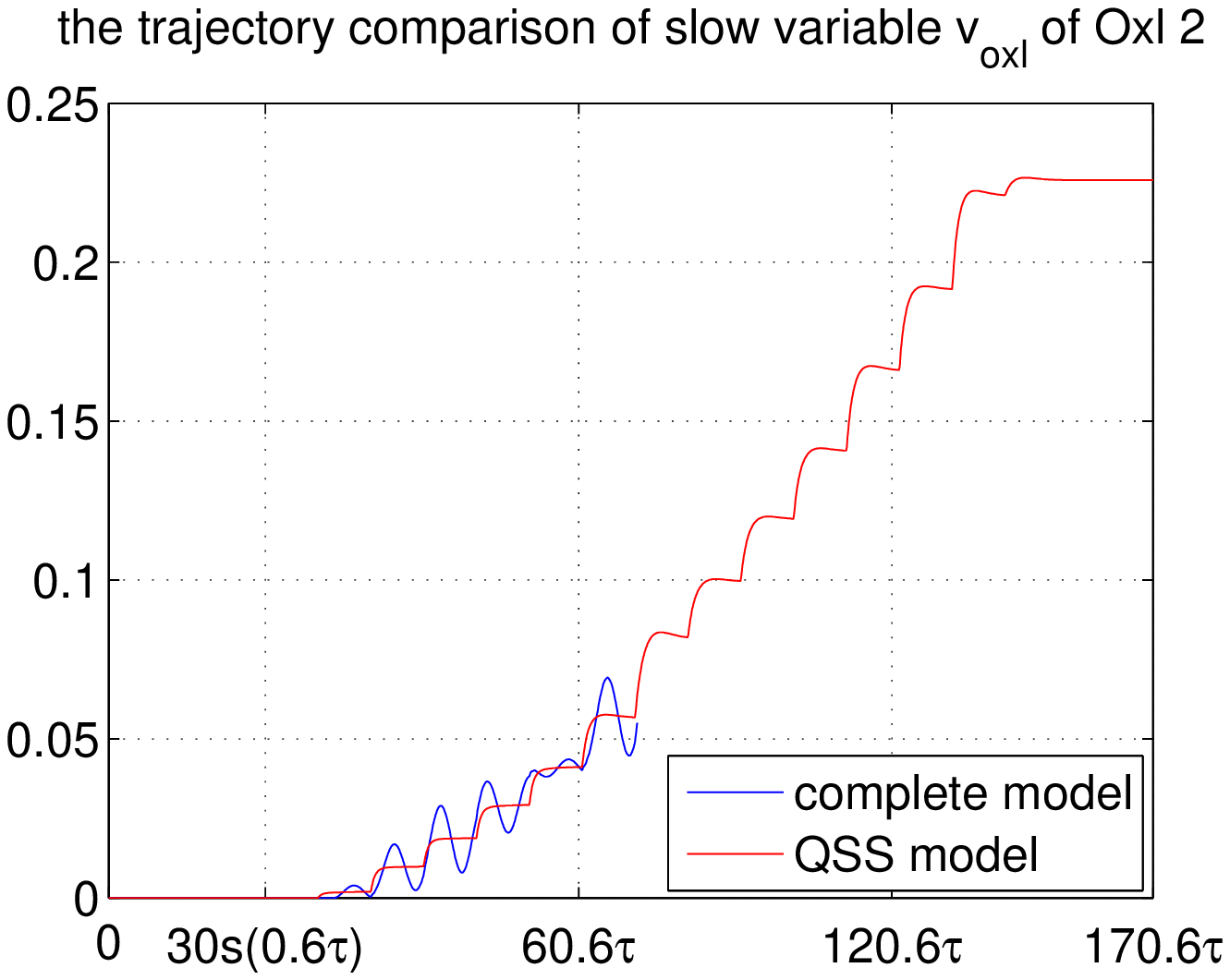}
\end{minipage}%
\begin{minipage}[t]{0.5\linewidth}
\includegraphics[width=1.8in ,keepaspectratio=true,angle=0]{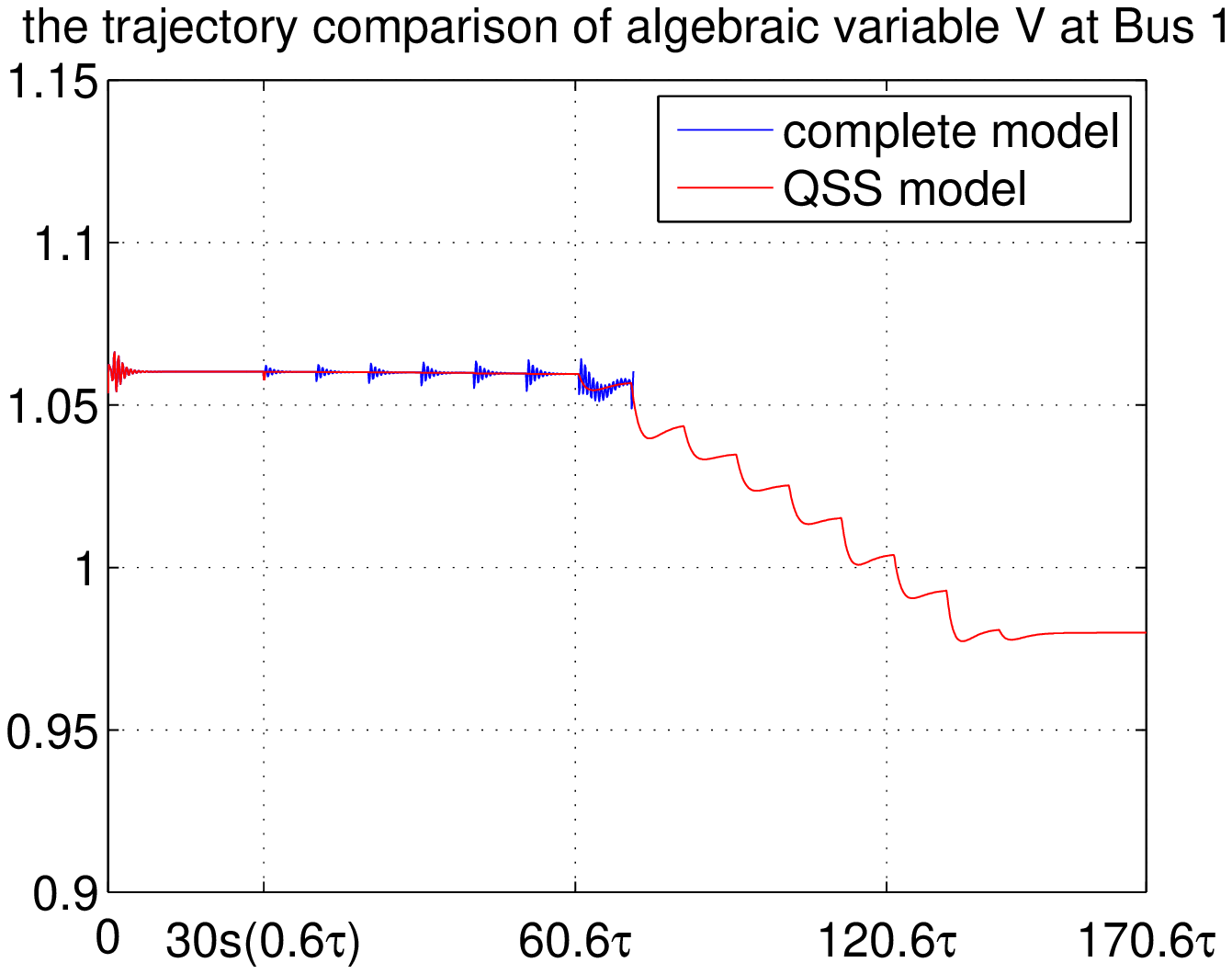}
\end{minipage}
\caption{The trajectory comparisons of the complete model and the QSS model for different variables. The assumption of QSS model that the fast variables are stable is not satisfied such that it gives incorrect approximations.}\label{my14completeqss_try}
\end{figure}

This system was set up based on the modified IEEE-14 bus system in Section \ref{sectionqssmodel}. Apart from the two turbine governors at Bus 1 and Bus 2 , there were three exponential recovery loads at Bus 9, Bus 10 and Bus 14 respectively, and five over excitation limiters were added for each exciter which started to work after a fixed delay 10s. Besides there were three load tap changers which are discrete models\cite{Cutsem:book}:
\begin{equation}\label{Ldeqn1}
m_{k+1}=\left\{\begin{array}{ll}m_k+\triangle{m}&\mbox{if  }v>v_0+d\mbox{  and  }m_k<m^{max}\\
m_k-\triangle{m}&\mbox{if  }v<v_0+d\mbox{  and  }m_k>m^{min}\\
m_k &\mbox{otherwise}\end{array}\right.
\end{equation}
where $m$ denotes the lap changer ratio. The one-line diagram of the modified system is also attached in Appendix \ref{numerical_ap}. There were two faults at Bus 9 and Bus 6 that happened simultaneously at 0.02s, and the faults were cleared by opening the breakers between Bus 7 and Bus 9, between Bus 6 and Bus 11 at 0.1s, and the one between Bus 6 and Bus 13 at 1s. The complete model was employed for the first 30s while the QSS model was employed afterwards. The comparison of trajectories of different variables in the complete model and the QSS model is showed in Fig. \ref{my14completeqss_try}.

In this case, the QSS model failed to give a correct approximation of the complete model. The time domain simulation of the complete model stopped and stated that there was "singularity likely" in the system around 101.2155s (71.8155$\tau$), while the QSS model did not encounter such problems and continued to converge to the long-term SEP. From Fig. \ref{my14completeqss_try}, it can be seen that in the complete model, fast dynamics $x$ were excited when slow variables evolved. The violent variation of fast variables $x$ due to slow variables finally resulted in voltage instability of the complete model such that it did not converge to the same asymptotically SEP as the QSS model. However, if we only look at the QSS model, the dynamics of fast variables $x$ due to slow variables are not noticeable since $x$ and $y$ converged to the transient SEPs immediately. Therefore, if the state of long-term SEP is acceptable, the post-fault system will be misclassified as stable. In this case, the assumption behind the QSS model that transient dynamics are stable in long-term time scale is violated.

\begin{figure}[!ht]
\begin{minipage}[t]{0.55\linewidth}
\includegraphics[width=1.80in ,keepaspectratio=true,angle=0]{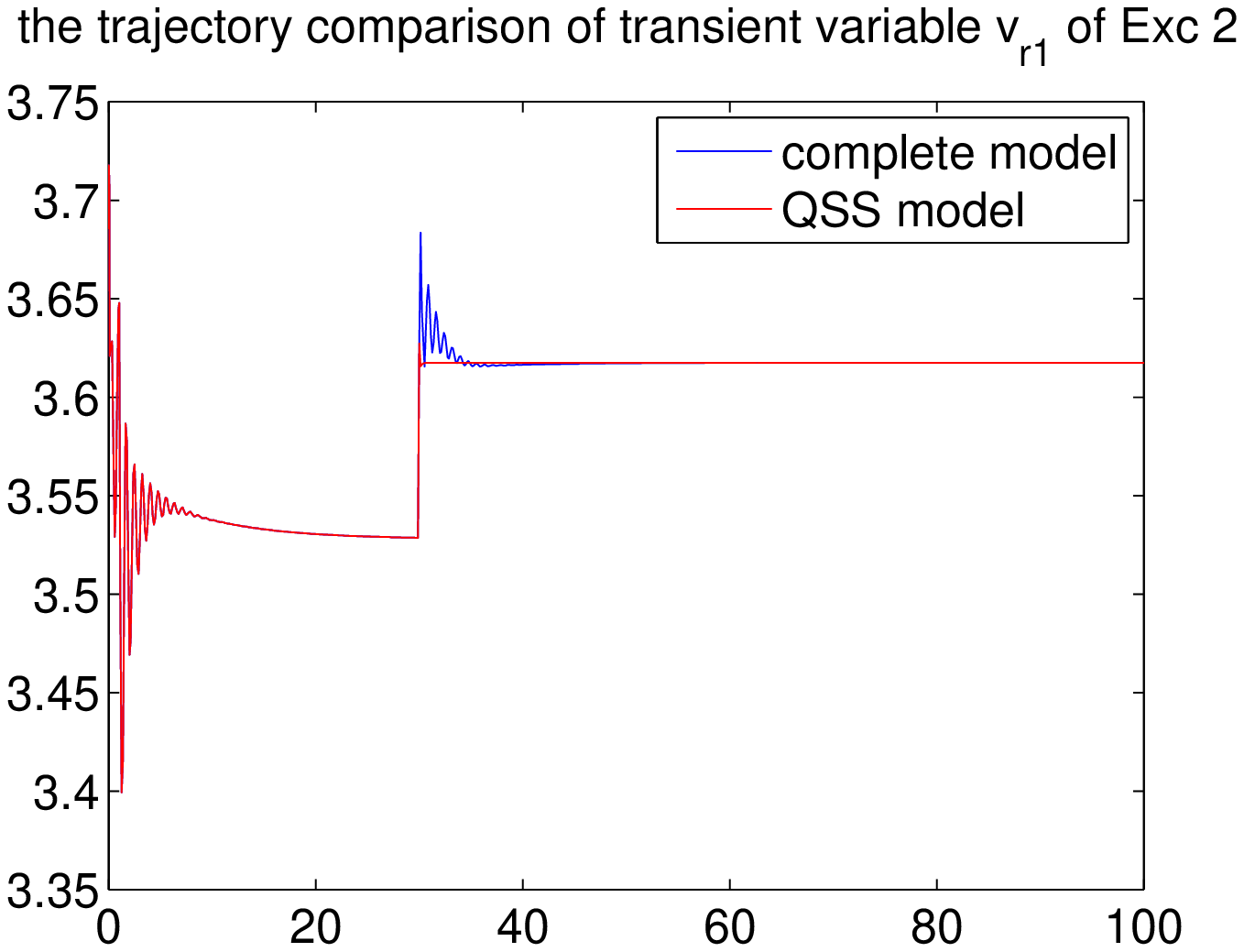}
\end{minipage}%
\begin{minipage}[t]{0.5\linewidth}
\includegraphics[width=1.8in,keepaspectratio=true,angle=0]{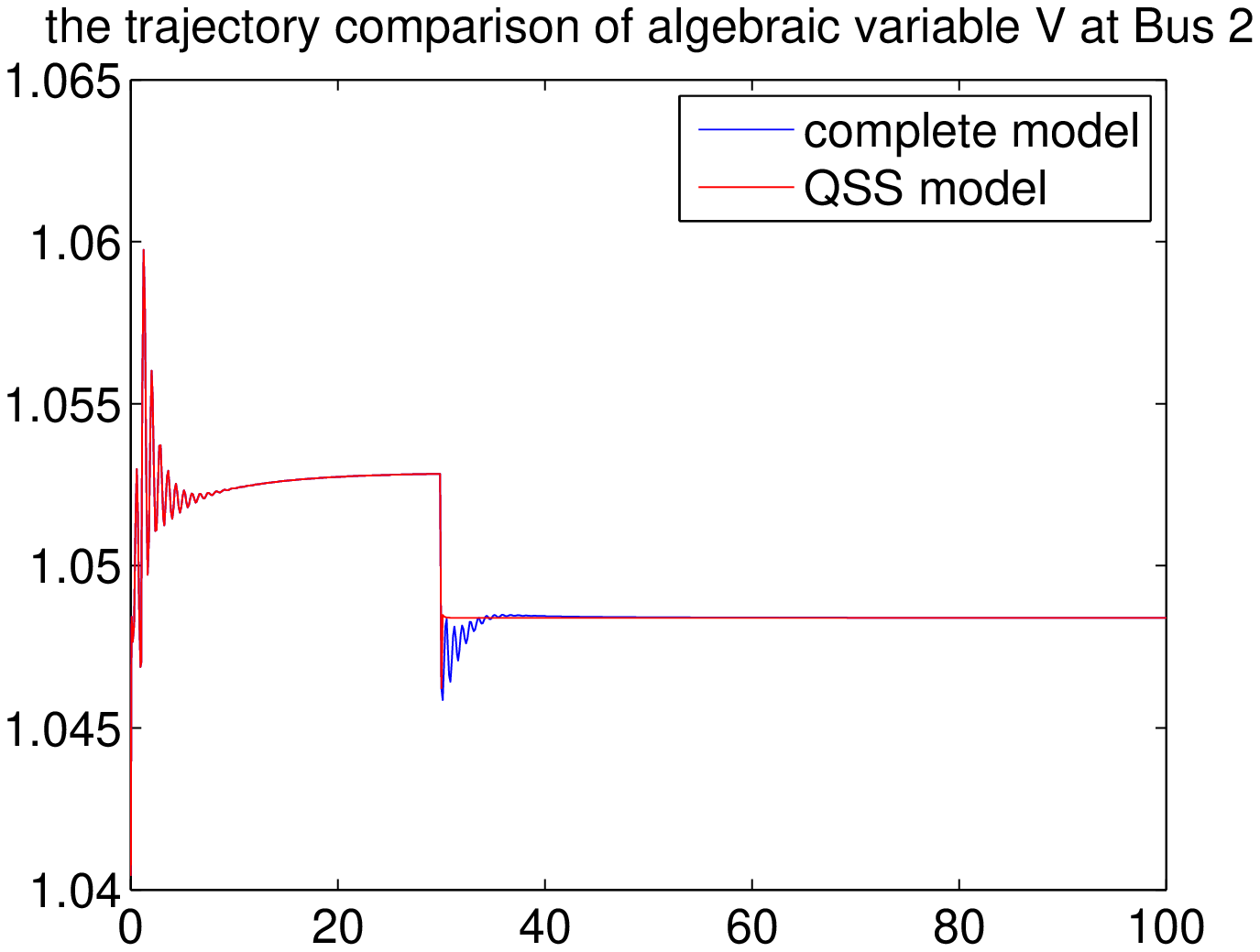}
\end{minipage}
\caption{The trajectories comparisons of the complete model and the QSS model for different variables when load tap changers changed at 30s. Both the complete model and the QSS model converged to the same SEP.}\label{my14try_31}
\end{figure}

\begin{figure}[!ht]
\begin{minipage}[t]{0.55\linewidth}
\includegraphics[width=1.8in,keepaspectratio=true,angle=0]{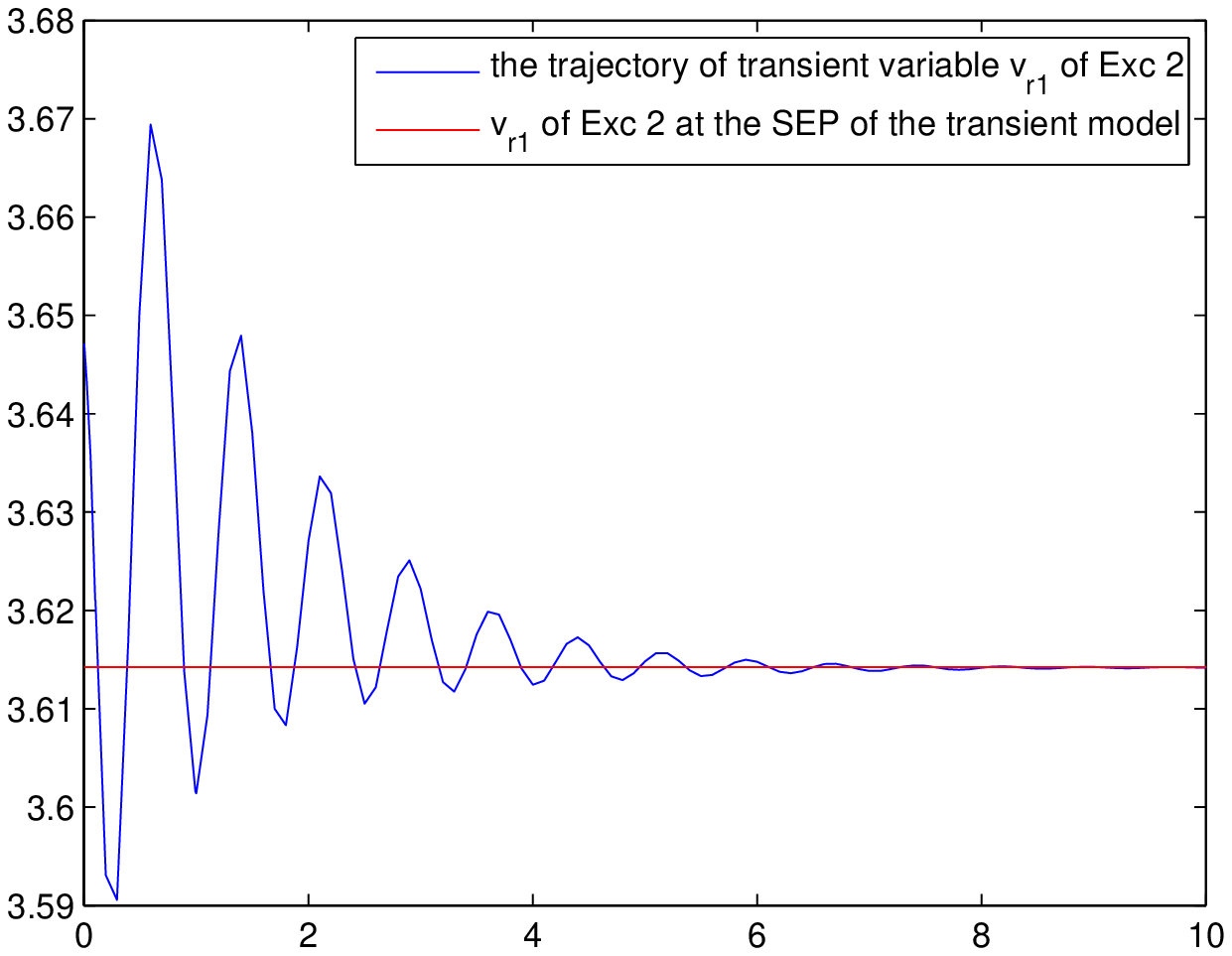}
\end{minipage}%
\begin{minipage}[t]{0.5\linewidth}
\includegraphics[width=1.8in,keepaspectratio=true,angle=0]{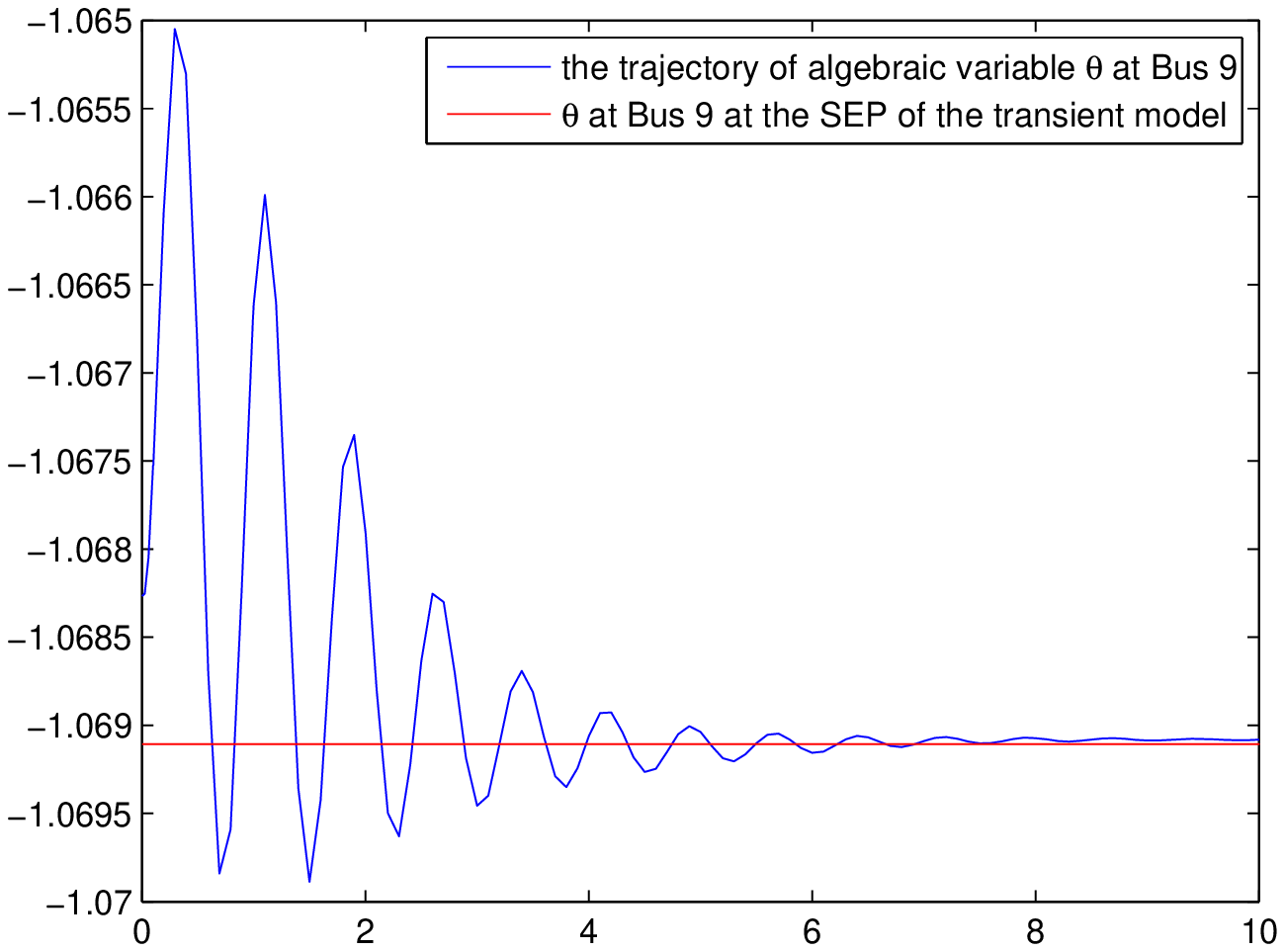}
\end{minipage}
\caption{The trajectories of the transient model when load tap changers changed at 30s which indicated that $(z_c^\star,z_d(2),x_0,y_0)$ was inside the stability region of the transient model.}\label{my14trytransient_31}
\end{figure}

The failure of the QSS model can be further explained by checking the trajectory of the transient model. When $z_d$ firstly changed from $z_d(1)$ to $z_d(2)$ at 30s, denote the initial point on the trajectory of the complete model when this change happened as $(z_c^\star,z_d(2),x_0,y_0)$, then the complete model fixed at $z_d(2)$ starting from $(z_c^\star,z_d(2),x_0,y_0)$:
\begin{eqnarray}\label{complete 2}
{z}_{c}^\prime&=&{h}_c({z_c,z_d(2),x,y})\\
\ee{x}^\prime&=&{f}({z_c,z_d(2),x,y})\nonumber\\
{0}&=&{g}({z_c,z_d(2),x,y})\nonumber
\end{eqnarray}
was stable as shown in Fig. (\ref{my14try_31}) in which both the complete model and the QSS model converged to the same long-term SEP. Moreover, the trajectories of two variables in the corresponding transient model starting from $(z_c^\star,z_d(2),x_0,y_0)$:
\begin{eqnarray}\label{qss 2}
\dot{x}&=&{f}({z_c,z_d(2),x,y})\\
{0}&=&{g}({z_c,z_d(2),x,y})\nonumber
\end{eqnarray}
is plotted in Fig. (\ref{my14trytransient_31}). It can be seen that both the fast variable and the algebraic variable converged to the SEP of the transient model (\ref{qss 2}). In other words, the initial point $(z_c^\star,z_d(2),x_0,y_0)$ of the complete model (\ref{complete 2}) is inside the stability region $A_t(z_c^{\star},z_d(2),x_{ts},y_{ts})$ of the transient model (\ref{qss 2}).

\begin{figure}[!ht]
\begin{minipage}[t]{0.55\linewidth}
\includegraphics[width=1.8in,keepaspectratio=true,angle=0]{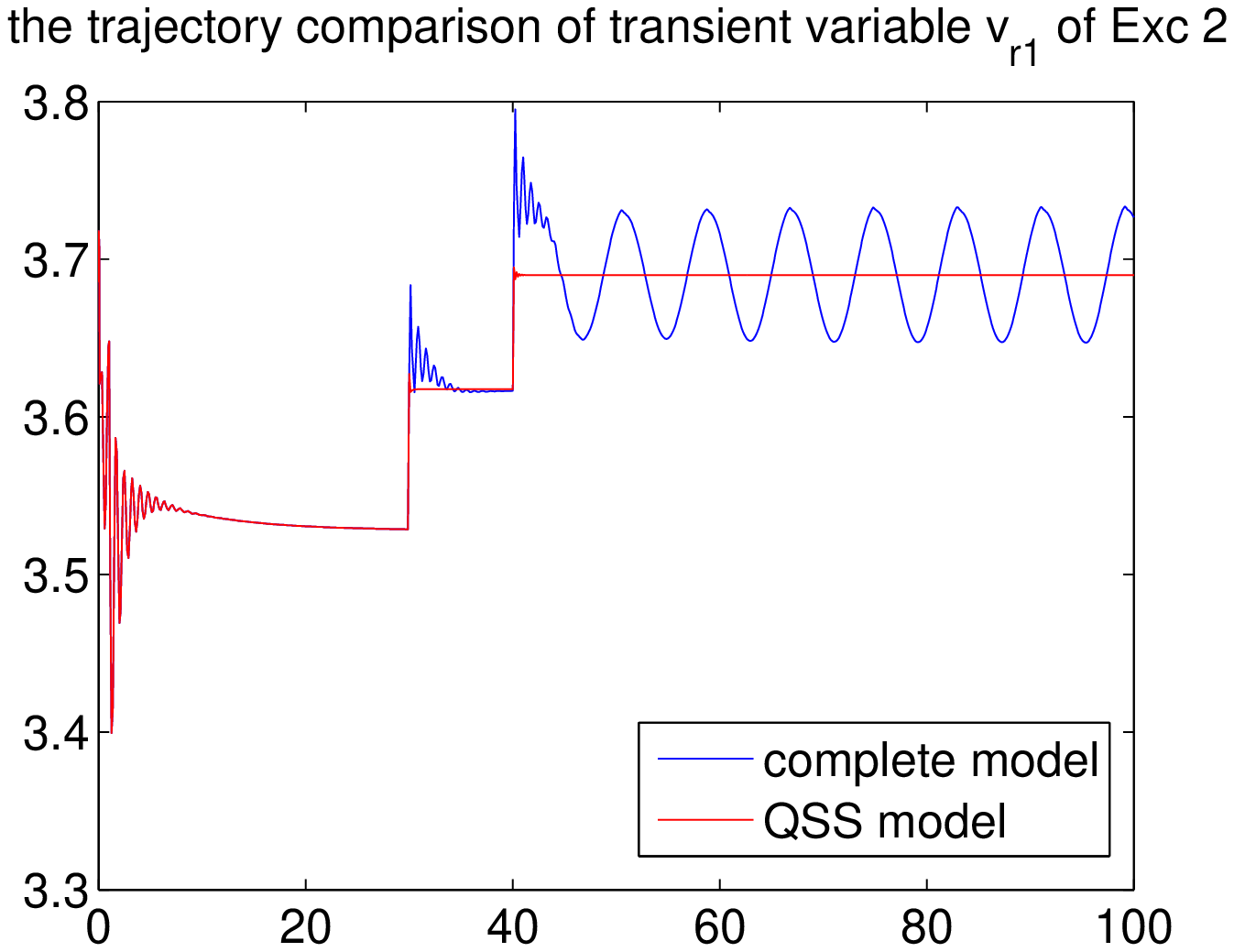}
\end{minipage}%
\begin{minipage}[t]{0.5\linewidth}
\includegraphics[width=1.8in,keepaspectratio=true,angle=0]{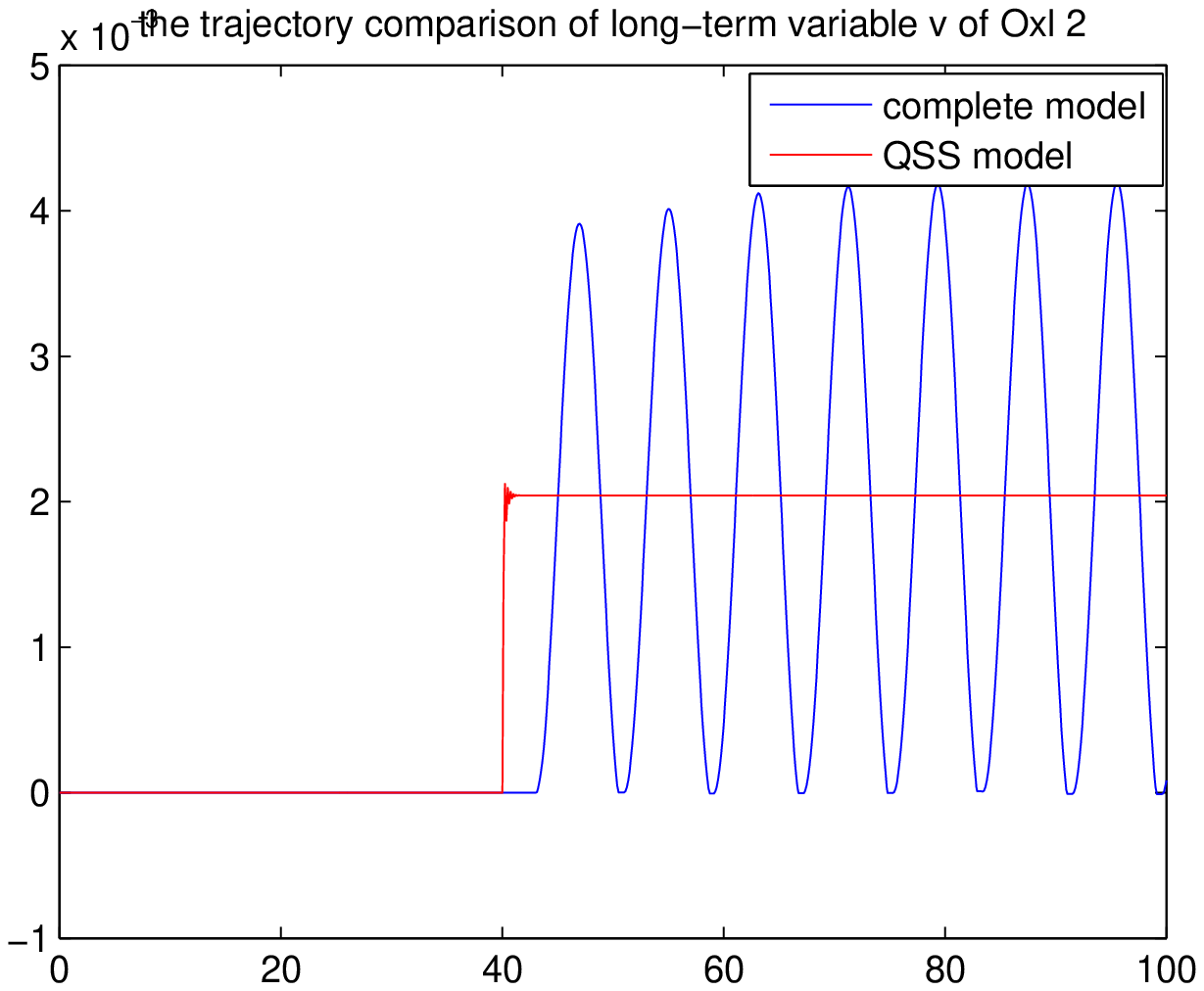}
\end{minipage}
\caption{The trajectories comparisons of the complete model and the QSS model for different variables when load tap changers changed at 40s. The complete model was unstable while the QSS model converged to a SEP.}\label{my14try_41}
\end{figure}

\begin{figure}[!ht]
\begin{minipage}[t]{0.55\linewidth}
\includegraphics[width=1.8in,keepaspectratio=true,angle=0]{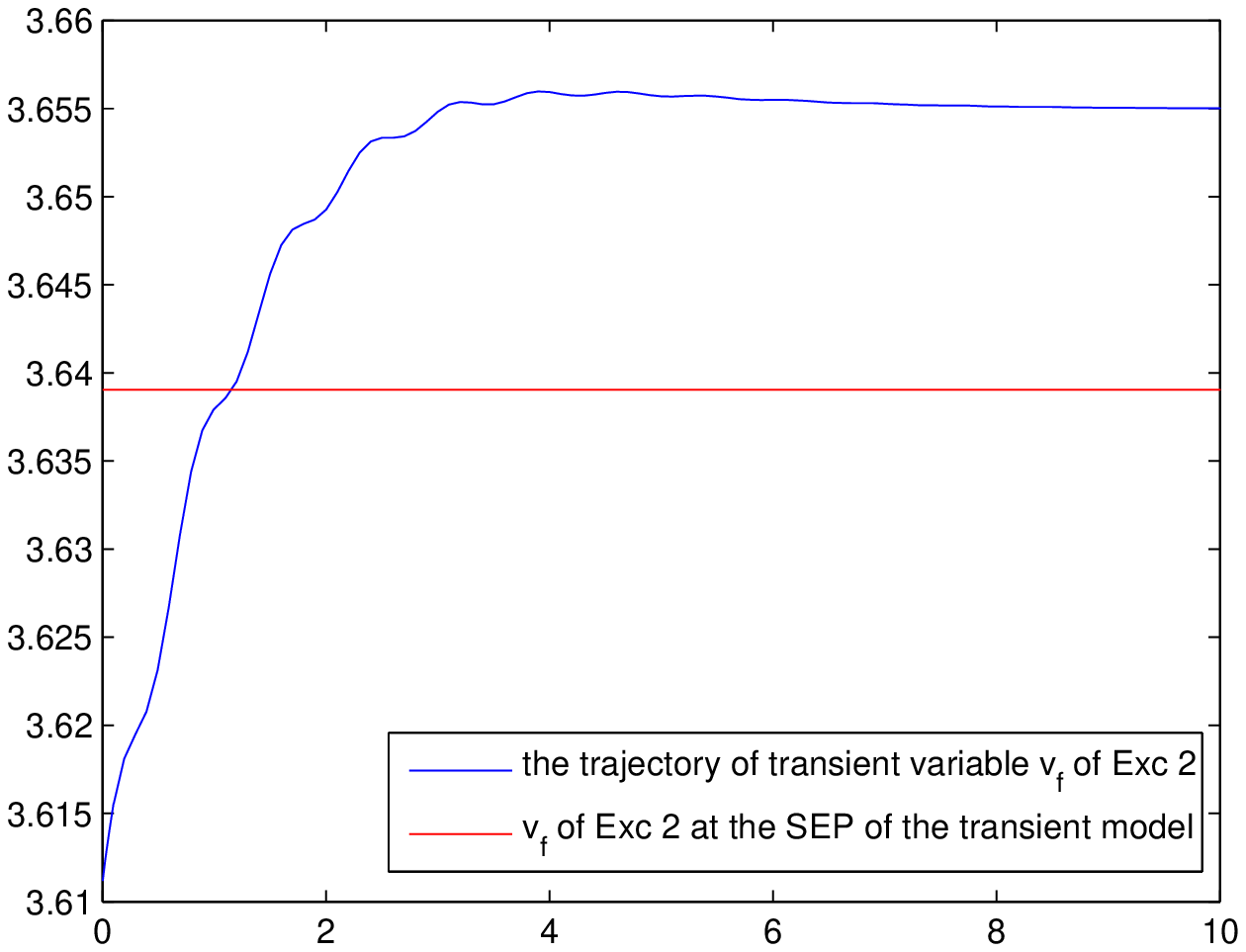}
\end{minipage}%
\begin{minipage}[t]{0.5\linewidth}
\includegraphics[width=1.8in ,keepaspectratio=true,angle=0]{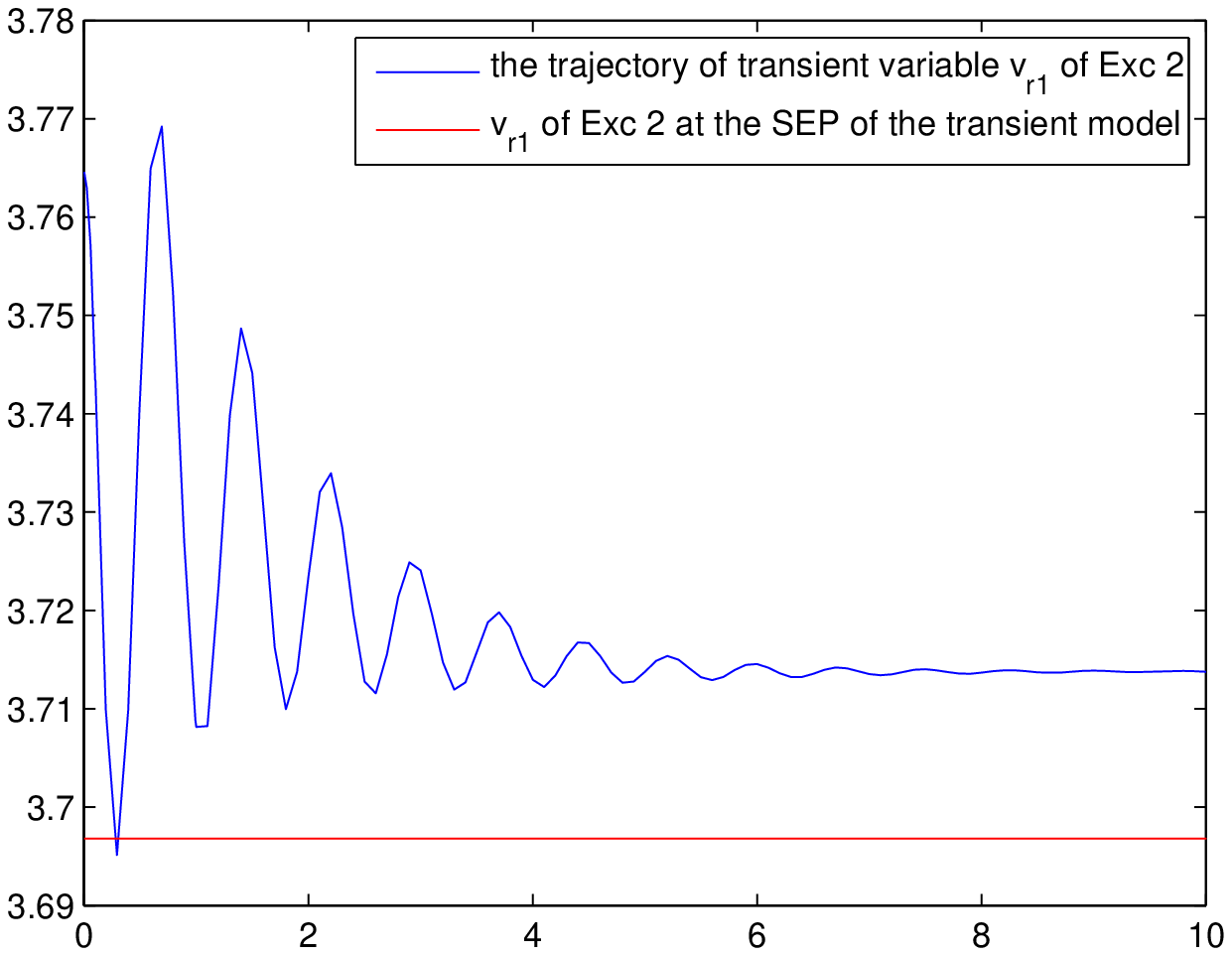}
\end{minipage}
\caption{The trajectories of the transient model when load tap changers changed at 40s which indicated that $(z_c^{\star\star},z_d(3),x_0,y_0)$ was outside of the stability region of the transient model.}\label{my14trytransient_41}
\end{figure}

However when $z_d$ changed from $z_d(2)$ to $z_d(3)$ at 40s, the complete model was no longer stable which can be seen from Fig. (\ref{my14try_41}). The fast variables were excited by the evolution of slow variables $z_d$ and $z_c$. The trajectories of fast variables in the corresponding transient model are plotted in Fig (\ref{my14trytransient_41}), and the initial point $(z_c^{\star\star},z_d(3),x_0,y_0)$ of the complete model (\ref{complete 2}) (substitute $z_d(2)$ by $z_d(3)$)
was outside of the stability region $A_t(z_c^{\star\star},z_d(3),x_{ts},y_{ts})$ of the transient model (\ref{qss 2}) (substitute $z_d(2)$ by $z_d(3)$)
As a result, the QSS model gives incorrect approximations of the complete model from then on.


\subsection{Numerical Example II}\label{qss_numerical 145}
\begin{figure}[!ht]
\begin{minipage}[t]{0.5\linewidth}
\includegraphics[width=1.75in,keepaspectratio=true,angle=0]{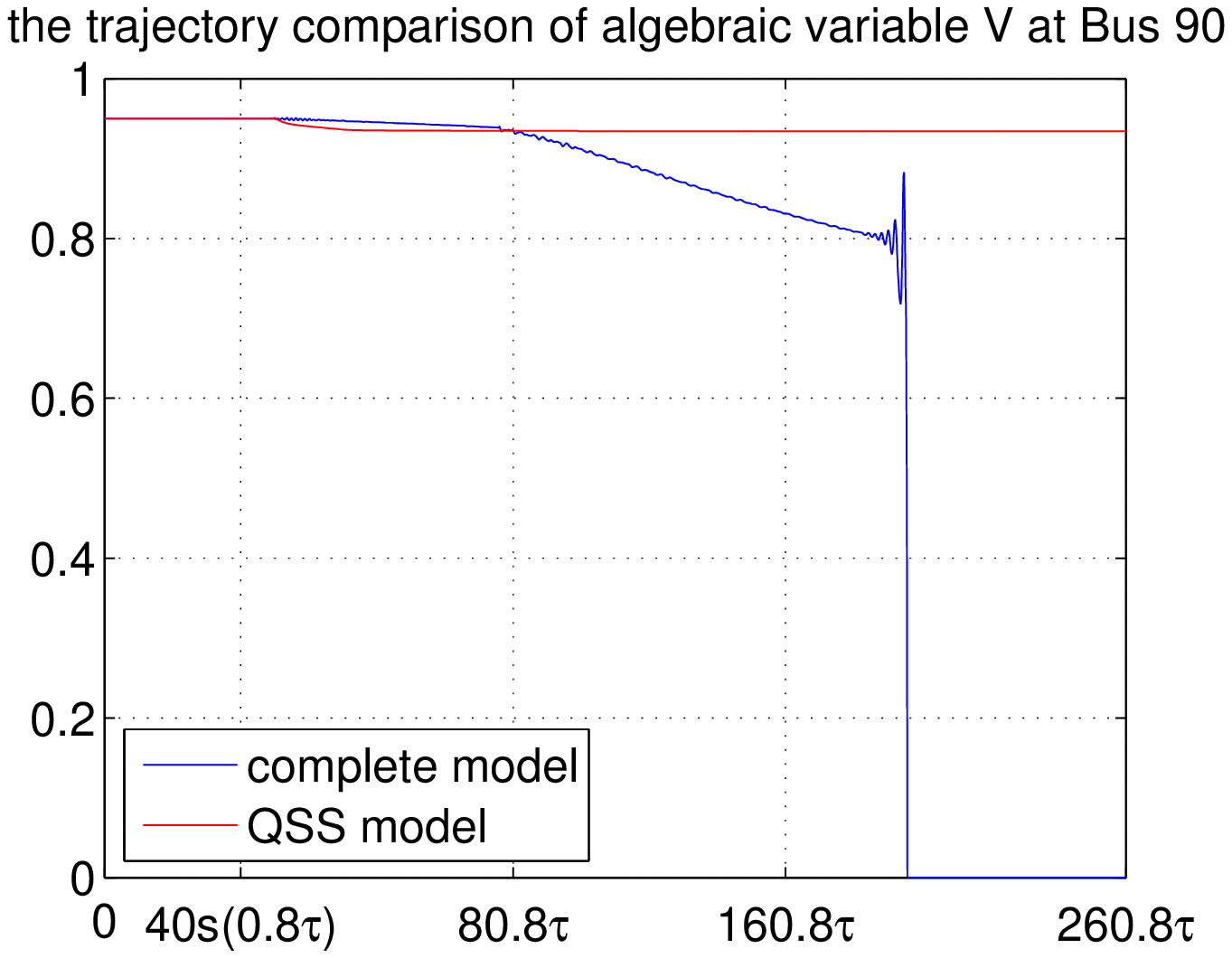}
\end{minipage}%
\begin{minipage}[t]{0.5\linewidth}
\includegraphics[width=1.75in,keepaspectratio=true,angle=0]{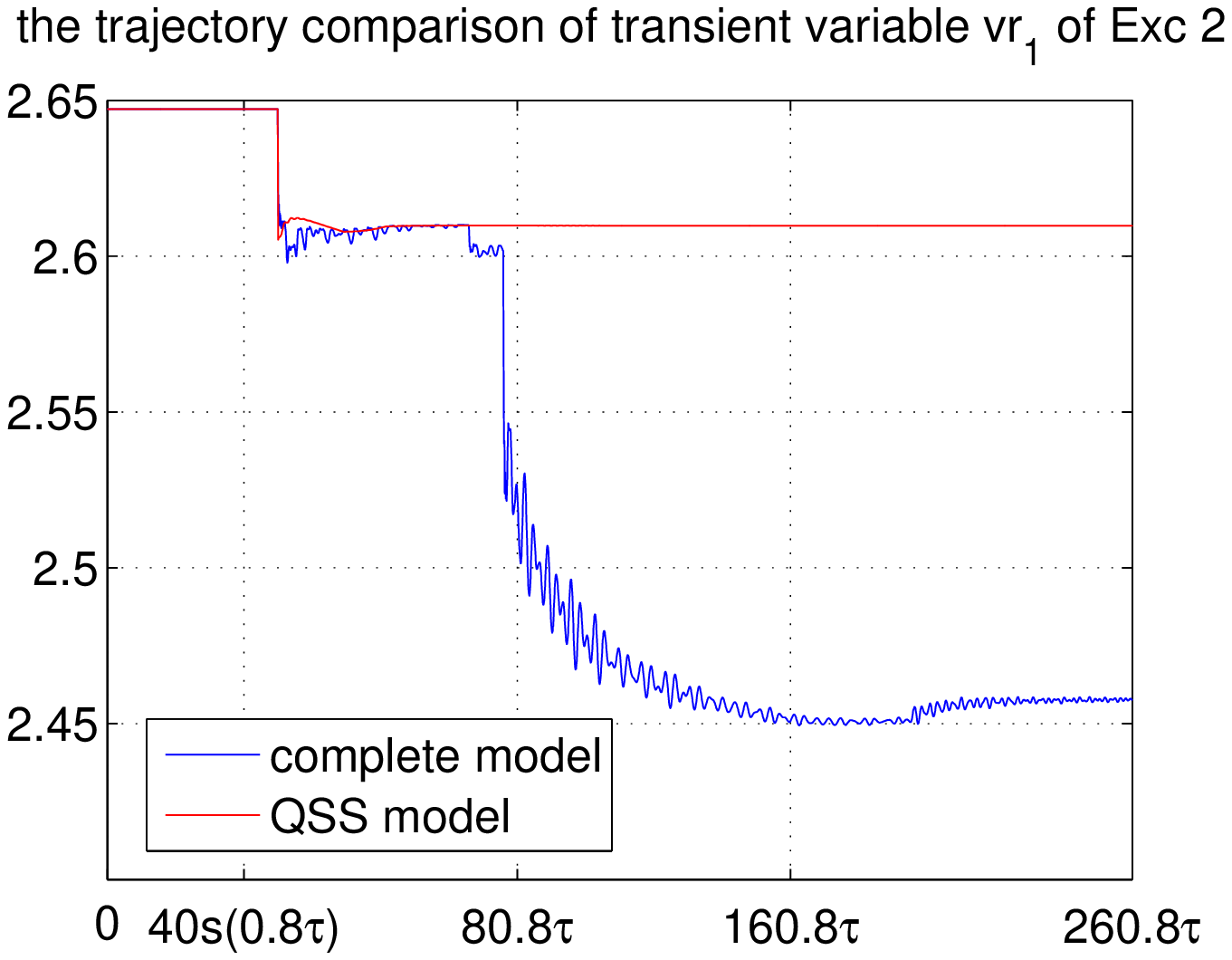}
\end{minipage}
\caption{The trajectory comparisons of the complete model and the QSS model. The QSS model converged to a long-term SEP while the complete model suffered from voltage collapse.}\label{my145}
\end{figure}

Another numerical example performed on a modified IEEE 145-bus system is presented below. Due to limited pages, only simulation results are shown in Fig. \ref{my145}. We can see that the voltage at Bus 90 was collapsed around 235s in the complete model, however, the voltage at Bus 90 settled down to the value around 0.9344 p.u in the QSS model. Also, the QSS model did not provide correct approximations for transient variables.

\section{Conclusion}\label{conclusion}
The QSS model was derived based on time-scale decomposition and it offers a good compromise between accuracy and efficiency. In this paper, two counter examples in which the QSS model provides inaccurate stability assessments are presented, and the reasons for the inability of the QSS model to approximate the complete model are explained from the stability regions of the transient models of the complete model. These counter examples suggest that there is a necessity to provide a theoretical foundation for the QSS model. Moreover, an improved QSS model may be needed in order to give consistently accurate approximation of the complete model.

\appendices
\section{The One-Line Diagram of Numerical Examples}\label{numerical_ap}

\begin{figure}[!ht]
\begin{minipage}[l]{0.5\linewidth}
\centering
\includegraphics[width=2in,keepaspectratio=true,angle=90]{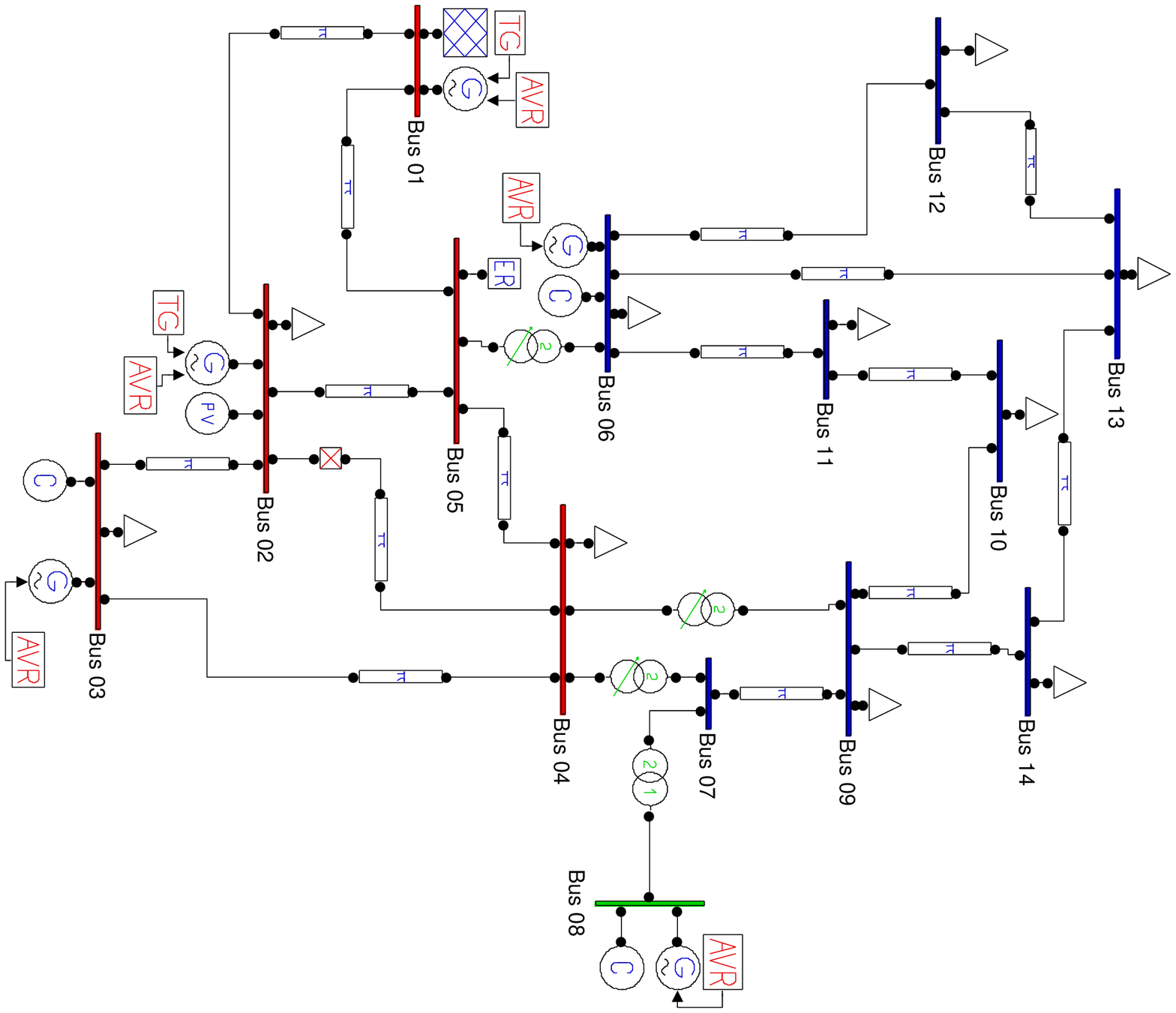}
\end{minipage}%
\begin{minipage}[c]{0.5\linewidth}
\centering
\includegraphics[width=2in,keepaspectratio=true,angle=-90]{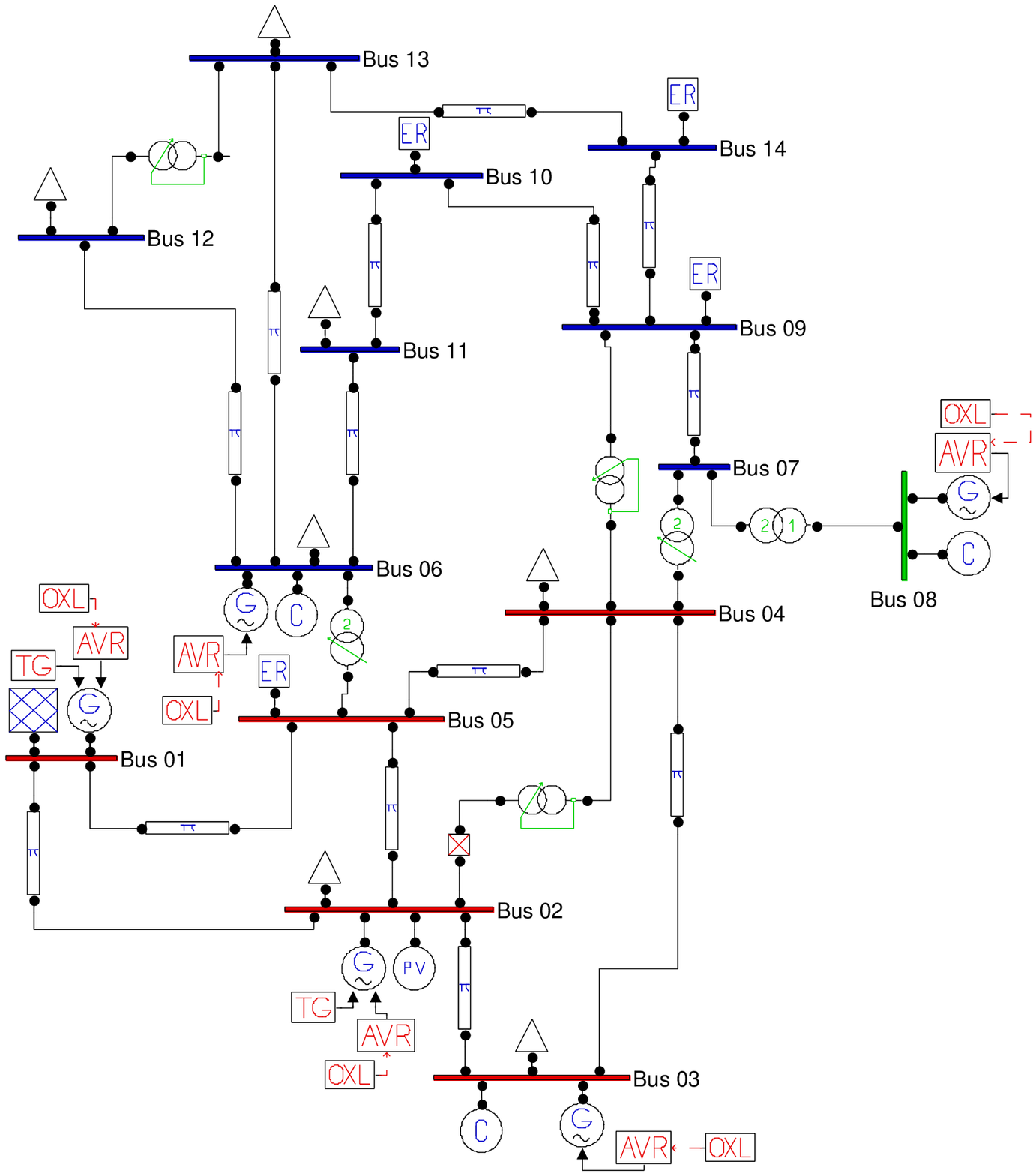}
\end{minipage}
\caption{One-line diagram of the example in Section {\ref{sectionqssmodel}}; One-line diagram of the example in Section {\ref{qss_numerical}}} \label{onelinediagram}
\end{figure}

The one-line diagram of the numerical examples are shown in Fig. \ref{onelinediagram}.

\section*{Acknowledgment}

The authors would like to thank Dr. Luis F. C. Alberto for helpful discussions. And this work was partially supported by the CERT through the National Energy Technology Laboratory Cooperative Agreement No. DE-FC26-09NT43321.

\ifCLASSOPTIONcaptionsoff
  \newpage
\fi

\end{document}